\documentclass{PoS}

\title{Fast multiwavelength variability from jets in X-ray binaries}

\ShortTitle{Fast multiwavelength variability from jets in X-ray binaries}

\author{\speaker{Piergiorgio Casella} , Thomas J. Maccarone\\
        School of Physics and Astronomy,
University of Southampton,
Southampton,
SO17 1BJ,
UK\\
        E-mail: \email{p.casella@soton.ac.uk}}
\author{Kieran O'Brien\\
Department of Physics,
University of California
Santa Barbara,
CA 93106, USA
}

\abstract{

While jets appear as a fundamental result of the accretion process onto compact objects in X-ray binaries, there is as yet no standard model for their underlying physics. The origin of the observed accretion disk-jet coupling also remain largely unknown. X-ray variability studies have revealed complex variability in the accretion flow onto stellar-mass black holes and neutron stars, on timescales as short as milliseconds. The detection of correlated broad-band rapid time variability in the jet emission would provide valuable information on how the variability is transferred along the jet and on the timescales of physical processes operating in these jets, ultimately helping to constrain internal jet physics, probe disk-jet coupling and infer accretion geometry. In recent years there have been indirect evidences for optical fast variability arising from a powerful jet. However, in optical and ultraviolet light the emission from the outer accretion disk can strongly contaminate the jet signal, which results in an ambiguous identification of the source of the observed variability. On the other hand, at much longer wavelengths the variability will be smeared out in time due as it comes from far out in the jet. Infrared variability studies are thus ideal for looking at jet variability on the fastest possible timescales. Thanks to newly available detectors and fast-readout modes, fast infrared and mid-infrared photometry is now possible. This is opening a new exciting window to study the geometry and the Physics of relativistic jets and their connection with the accretion flow. Here we present the first results from a large ongoing fast-timing multi-wavelength project, showing the first unambiguous evidence for sub-second jet variable emission. We show how this type of data already allows us to put quantitative constraints to the jet speed, geometry and physics, and discuss the great potential of new observations in the near future.
}

\FullConference{High Time Resolution Astrophysics IV - The Era of Extremely Large Telescopes - HTRA-IV,\\
		May 5-7, 2010\\
		Agios Nikolaos, Crete, Greece}

\begin{document}

\section{Introduction}

To date, jets have been discovered in a variety of astronomical objects over a wide range of gravitational regimes, from planetary nebulae to accreting white dwarfs, from stellar-mass black holes to supermassive AGN. Clearly, to reach a good understanding of such a general phenomenon would be interesting in its own right, but it would also be important given the important role that such jets have on the evolution of the launching systems (because of the power carried away from the accreting system), and given the influence they have on the surrounding media (e.g., the energetic feedback of relativistic jets in supermassive black holes has cosmological consequences).

The wealth of multi-wavelength observations of X-ray binaries (XBs) over the past decade have made clear the ubiquity of jets in these systems \cite{fender06}. The study of XB jets can be an important tool to understand the physics of jets and their link with the accretion flow, as the luminosity (thus plausibly the accretion rate) in these systems shows variability over several order of magnitudes and on timescales ranging from milliseconds to decades.

\begin{figure} 
\begin{center}
\includegraphics[width=.8\textwidth]{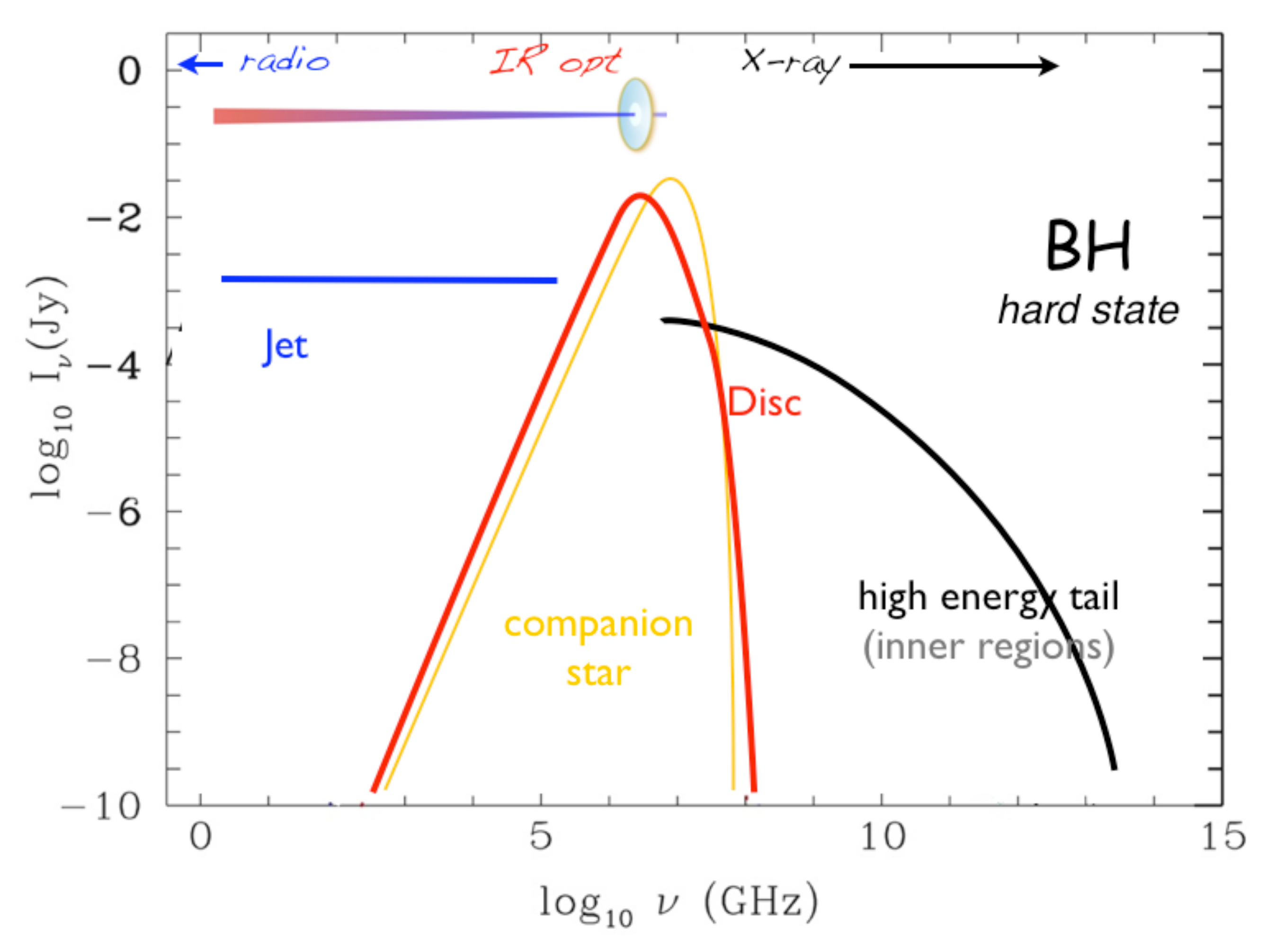}
\end{center}
\caption{Schematic representation of a typical broad-band energy spectrum of a black-hole X-ray binary in its hard state. The main spectral components are indicated and highly simplified. Several spectral components contribute to the optical emission, with different proportions depending on the source, while going toward longer wavelengths the jet becomes more and more dominant.}
\label{sed} 
\end{figure}

The presence of jets in XBs had been first proposed to explain the radio emission often observed in these objects and in analogy to the well known jets in AGN, and then confirmed in several cases though imaging. The overall spectrum of XBs is rather complex (see a simplified schematic view in Fig.~\ref{sed}): the soft X-ray flux is generally believed to come predominantly from an accretion disc around the compact object, whose emission can extend down to optical or infrared wavelengths, while the hard X-ray flux is thought to arise from a hot Comptonizing corona and/or from the jet. Recently it has been shown that also the infrared emission includes a substantial contribution from the relativistic jet, in the hard states of XBs \cite{corbelfender02,russelletal07}.  Despite the rapid increase in our phenomenological understanding of jets from XBs, we still lack a fundamental understanding of how jets are powered and collimated, or what the bulk and internal properties of the jets are.

High-speed simultaneous optical/X-ray photometry of three accreting black holes (BHs)
opened a new promising window. Complex correlated variability in the optical and X-ray emission
\cite{spruitkanbach02} was seen from XTE J1118+480, while fast optical photometry of SWIFT J1753.5-0127 \cite{durantetal08} and GX 339--4 \cite{gandhietal08} revealed further complexity. \cite{malzacetal04} explained the behaviour observed in XTE J1118+480 through coupling of an optically emitting jet and an X-ray emitting corona in a common energy reservoir. An alternative explanation comes from the magnetically driven disc corona model \cite{merlonietal00}: magnetic flares happen in an accretion disc corona where thermal cyclo-synchrotron emission contributes significantly to the optical emission, while the X-rays are produced by Comptonization of the soft photons produced by dissipation in the underlying disc and by the synchrotron process itself.  The two explanations differ substantially in the predictions at infrared wavelengths, where a jet appears as the most probable origin for the emission \cite[see also Fig.~\ref{sed}]{russelletal06}.

In order to solve these ambiguities and securely identify the jet variable component, we have started a large multi-wavelengh program, aimed at performing fast-timing simultaneous observations at different wavelengths of several XBs hosting both black holes and neutron stars. Here we report on the result of the first obtained dataset.

\section{The first dataset: ISAAC + RXTE observations of GX 339--4}

The BH candidate GX 339--4 is a recurrent X-ray transient \cite{markert73}. It has been detected as a highly variable source from radio through hard X-rays \cite[and references therein]{makishimaetal86,corbeletal00,coriatetal09}. Optical spectroscopy indicates a mass function of 5.8 $\pm$ 0.5 M$\odot$ and a minimum distance of 6 kpc \cite{hynesetal03,hynesetal04}.  It is the first BH XB for which fast optical/X-ray correlated variability was observed \cite{motchetal82}. Multiwavelength campaigns clearly reveal a non-thermal contribution to the infrared emission in the hard state, most probably arising from a compact jet \cite{corbelfender02}. Thus, we observed this source with high time resolution simultaneously in infrared and X-rays, aiming at identifying a possible variable jet component.

\subsection{Data reduction}

\noindent \underline{\it Infrared}

We observed GX339-4 from ESO's Paranal Observatory on 2008 August 18th.  We obtained fast Ks-band photometry with ISAAC (Moorwood et al. 1998) mounted on the 8.2-m UT1/ANTU telescope. The 23'' x 23'' window used encompassed the target, a bright 'reference' star located 13.6 arcsec south of our target and a fainter 'comparison' star  8.9-arcsec north-east of GX339-4 (respectively 2MASS17024972-4847361; Ks=9.5 and 2MASS17024995-4847161; Ks=12.8). 

We used the "FastJitter mode" with a time resolution of 62.5 ms. This generated cubes of data with 2500 images in each cube and a small deadtime between cubes. The ULTRACAM pipeline\footnote{We thank Tom Marsh for the use of the software (http://deneb.astro.warwick.ac.uk/phsaap/software/).} was used for the data reduction. We performed aperture photometry of the three sources (target, reference and comparison stars) and used the bright reference star for relative photometry of the target and comparison stars. The positions of the aperture regions around the target and the comparison star were linked to the position of the bright reference star to allow for image motion and were updated at each time step. The atmospheric conditions were good and the resulting light curve for the comparison star was consistent with a constant, as expected. By combining all 250000 images, we estimate an average magnitude of Ks=12.4$\pm$0.2 for GX 339--4, which corresponds to an average flux of F$\sim 10^{-11} erg s^{-1} cm^{-2}$ .

\begin{figure} 
\includegraphics[width=.505\textwidth]{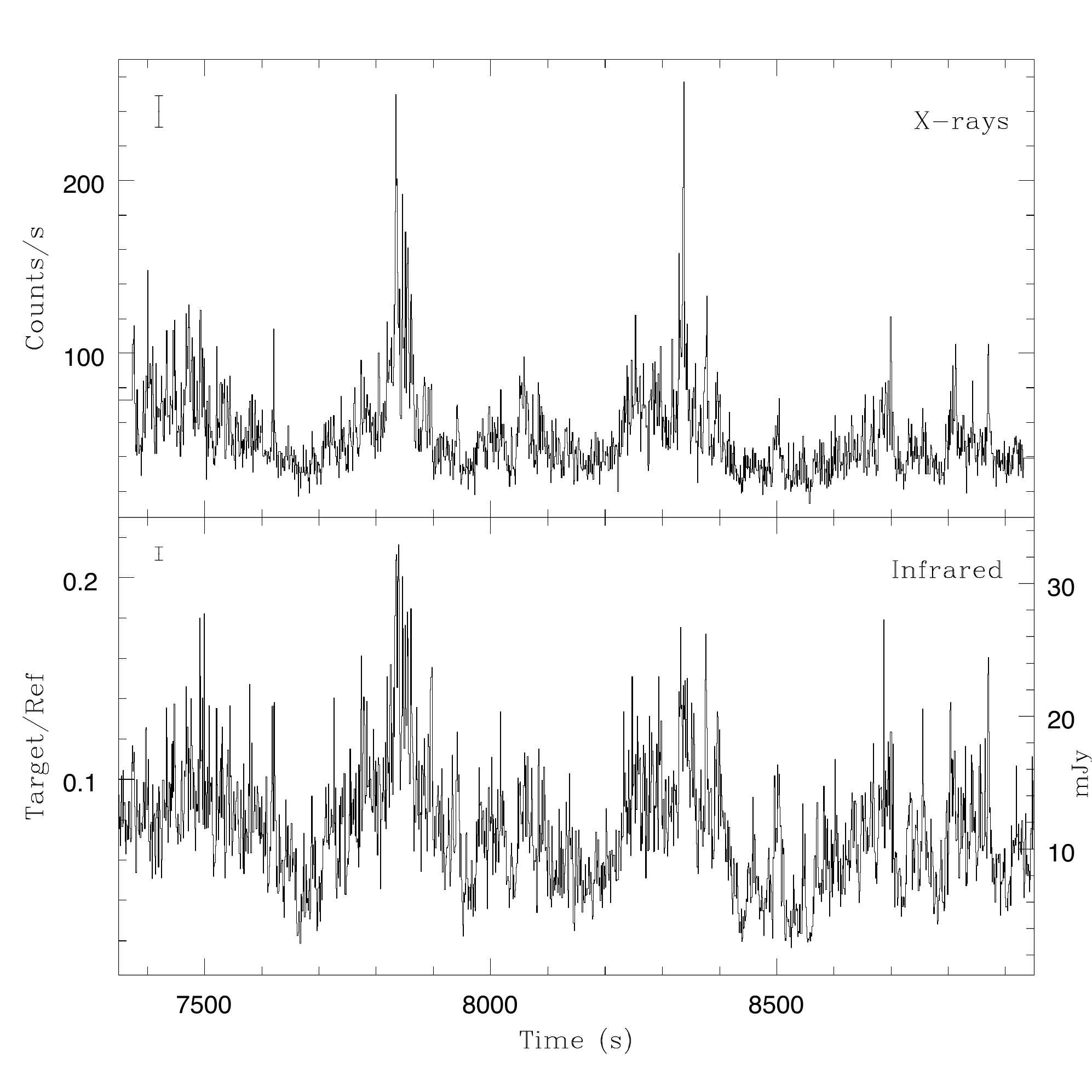} \includegraphics[width=.5\textwidth]{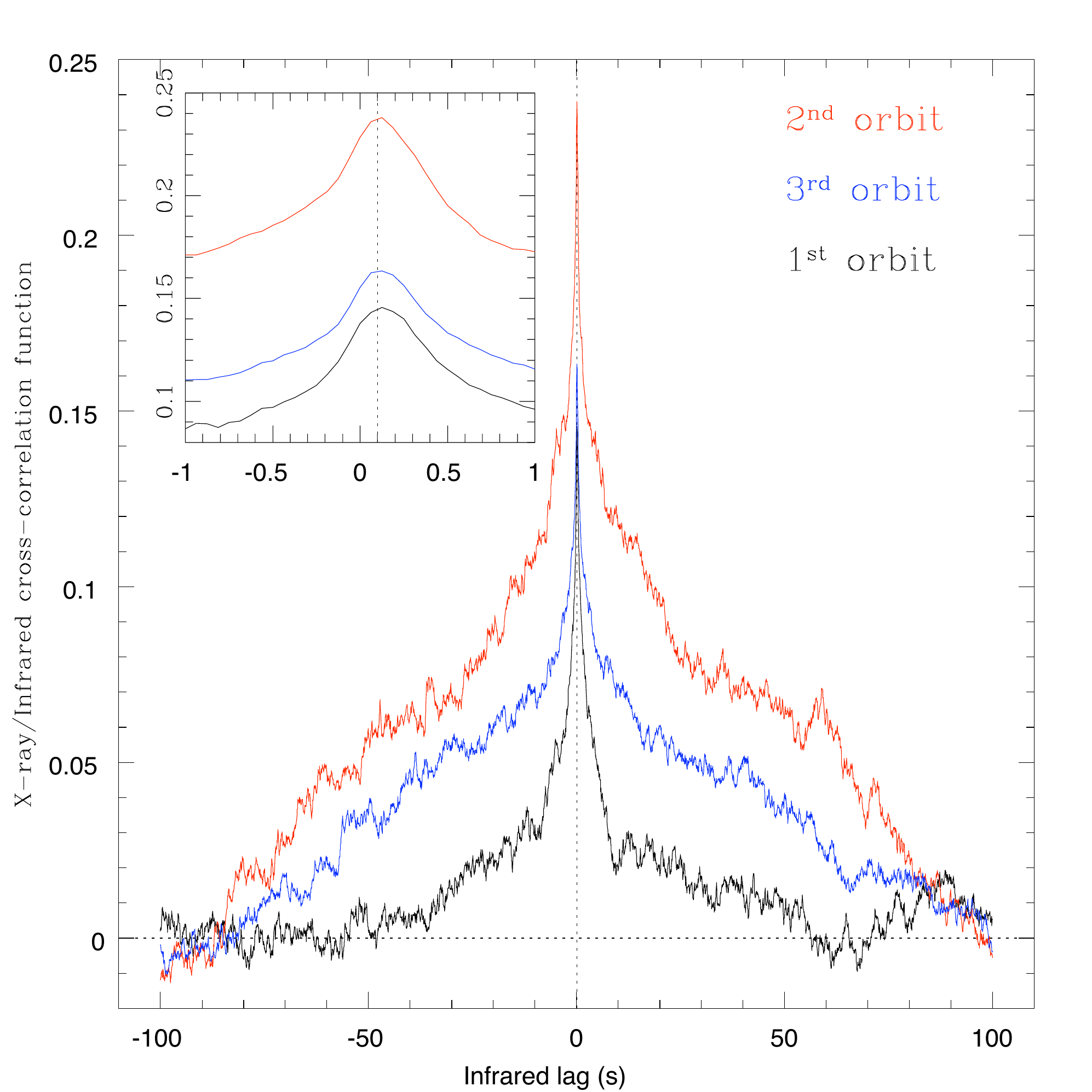}
\caption{{\it Left top panel:} A sample of the X-ray light curve of GX 339--4, obtained with the PCA onboard RXTE (2nd out of 3 satellite orbits). The data are background subtracted, in the 2-15 keV energy range, at 1-second time resolution. {\it Left bottom panel:} The simultaneous infrared light curve, obtained with ISAAC. We show the ratio between the source (average $4.4\times10^5$ counts/s) and the reference-star ($6\times10^6$ counts/s) count rates in the K$_S$ filter, at 1-second time resolution. The right ordinates show the de-reddened flux. We show the typical error bars in the top-left corner of each panel. {\it Right:} Cross-correlations of the X-ray and infrared light curves of GX 339--4 (positive lags mean infrared lags the X-rays). A strong, nearly symmetric correlation is evident in all the three time intervals, corresponding to different RXTE orbits. In the inset we show a zoom of the peaks, showing the infrared delay of $\sim$100~ms with respect to the X-rays. The inset also shows a slight asymmetry toward positive delays. } 
\label{lcccf} 
\end{figure}

The time-stamp was generated from the DATE-OBS fits keyword, which represents the start time of the first image and the exposure time (DIT) of each subsequent image. A sample of the highly variable light curve for GX 339--4 is shown in the left-bottom panel of Fig. \ref{lcccf}.

\noindent \underline{\it X-rays}

Simultaneously with the infrared observations, GX 339--4 was observed with
the Proportional Counter Array (PCA) onboard the {\it Rossi X-ray
Timing Explorer (RXTE)}. Two proportional counter units (PCUs) were
active during the whole observation. The X-ray data span three
consecutive satellite orbits, for a total exposure of 4.6 ksec.
The {\tt Binned Mode} (8~ms time resolution)
was used for this analysis, using the 2-15 keV energy range
(channels 0-35). The barycenter correction for Earth and satellite motion was
applied.  Standard HEADAS 6.5.1 tools were used for data reduction. In
the left-upper panel of Fig.~\ref{lcccf} we show a sample of the light
curve, corresponding to the second {\it RXTE} orbit. Spectral fitting
with a power-law with photon index 1.6 results in a 2--10 keV unabsorbed flux of
$F_{\rm X}\sim1.4\times 10^{-10}$ erg~s$^{-1}$~cm$^{-2}$ (corresponding to a luminosity of $L_{\rm X}\sim6\times 10^{35} ({d \over 6~kpc})^2$ erg~s$^{-1}$).

\begin{figure} 
 \includegraphics[width=.485\textwidth]{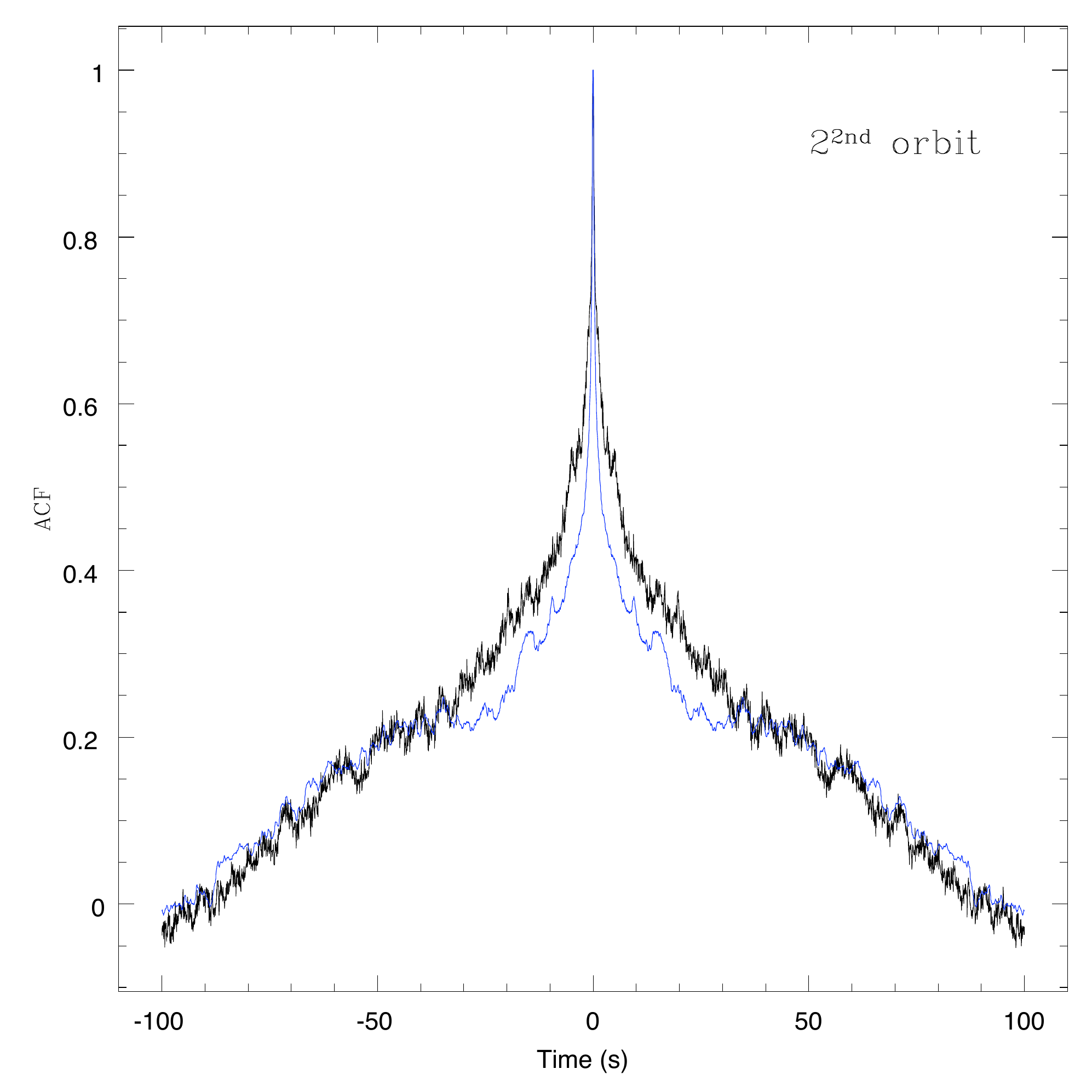} \includegraphics[width=.5\textwidth]{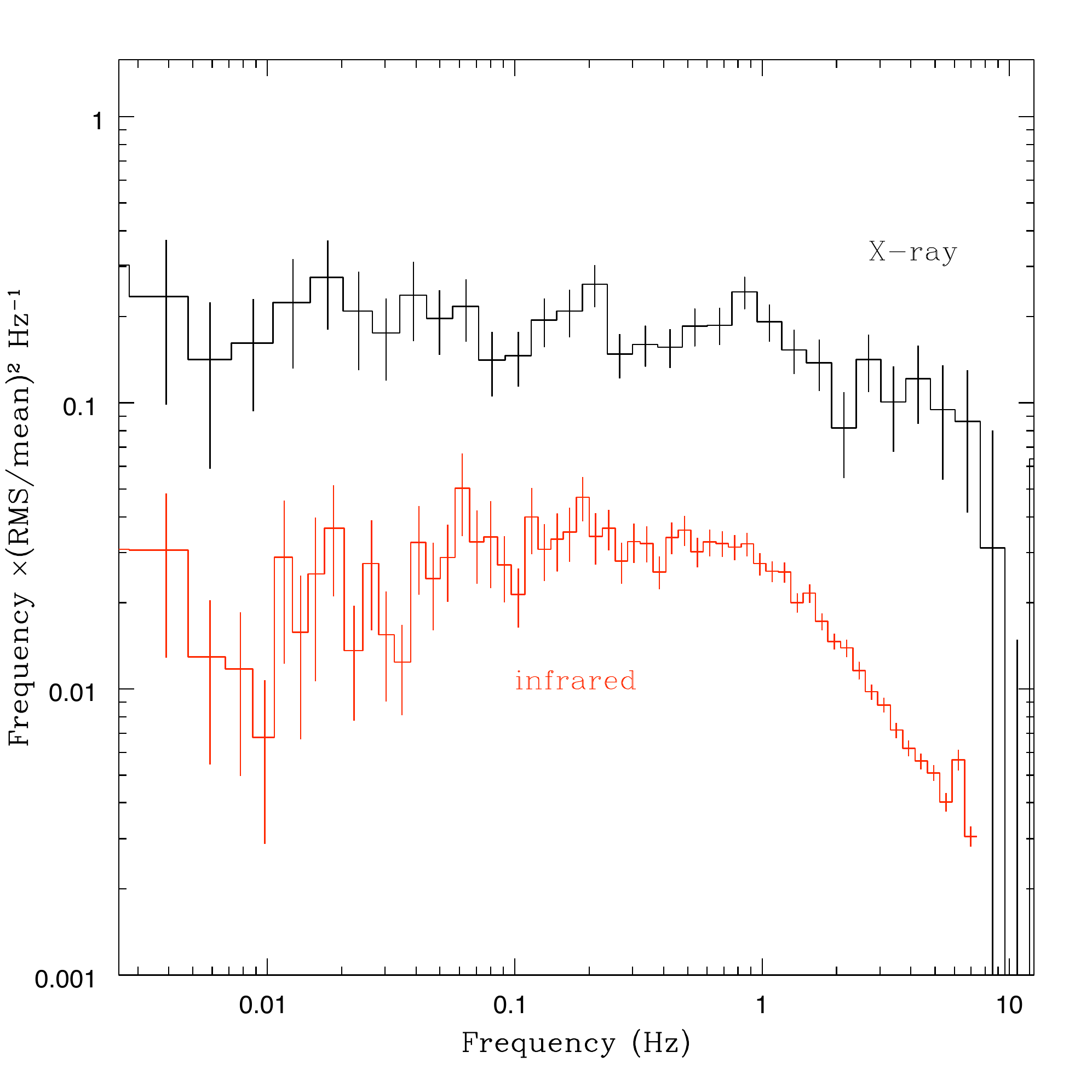} 
\caption{{\it Left:} Auto-correlations of the X-ray and infrared light curves of GX 339--4 for the 2nd RXTE orbit. The two auto-correlation functions are somewhat overlapping on timescales of $\sim$40 seconds or longer, but the optical one becomes clearly narrower on short timescales. {\it Right:} X-ray (2-15 keV) power spectrum of the second RXTE orbit (upper curve), together with the power spectrum of the simultaneous infrared light curve (lower curve). The Poissonian noise has been subtracted from both spectra. The peak at $\sim$6 Hz in the infrared spectrum is instrumental. The high-frequency portion of the infrared spectrum has yet un-modeled systematics, which however do not affect the results presented here.}
\label{acfpds} 
\end{figure}

\subsection{Timing analysis}

Both datasets have an absolute time accuracy better than the time resolution used here: ISAAC data have a timing accuracy of about 10 ms (the readout time), while RXTE data have a timing accuracy of 2.5 $\mu$s \cite{jahodaetal06}.

From the light curves shown in the left panels of Fig.~\ref{lcccf} a strong correlation between X-ray and infrared flux is evident. Both long, smooth variability and short, sharper flares appear with similar relative amplitude in the two energy bands. In order to measure any time delay, we calculated a cross-correlation function (CCF) for each of the three {\it RXTE} orbits, without applying any de-trending procedure. The results are shown in the right panel of Fig.~\ref{lcccf}. The strong correlation is confirmed. The CCF appears highly symmetric and relatively stable over the three time intervals, with the change in amplitude simply reflecting the different variability amplitude in the light curves themselves. In the inset, we show a zoom on the peak of the CCF, which shows how the infrared emission lags the X-rays by 0.1 seconds, to which we associate an uncertainty of 30$\%$ (which includes systematics).

For each RXTE orbit, we also calculated the auto-correlation functions and the Fourier power spectra of both the infrared and the X-ray light curves (in Fig.~\ref{acfpds} we plot those corresponding to the 2nd RXTE orbit). The Fourier power spectra were calculated after filling the gaps in the infrared light curve with simulated Poissonian noise. Different filling methods do not change the resulting power spectra significantly, especially at high frequencies.

\subsection{Strong evidences for a flickering jet}

The main result of our work is the discovery of a strong correlation between the infrared and the X-ray variability in GX 339--4. The fact that the CCF is nearly symmetric and peaks at 100~ms, together with the optical ACFs being narrower than the X-ray ones (at least on timescales shorter than a few tens of seconds, see left panel of Fig.\ref{acfpds}) rules out a reprocessing origin for the infrared variability. If the infrared radiation arose from reprocessing of X-rays by the outer disk, the short time delay would imply a highly inclined disk. This would produce an highly asymmetric CCF, with a tail at long lags \cite{obrienetal02}.

Additionally, power spectral analysis shows significant infrared variability (at least 5\% fractional rms, see right panel of Fig.~\ref{acfpds}) on timescales of $\sim$200~ms or shorter, which sets an upper limit of $\sim 6 \times$ 10$^9$ cm to the radius of the infrared-emitting region. From (5\% of) the observed infrared average flux of $F\sim 1.5\times10^{-11}$ erg s$^{-1}$ cm$^{-2}$, we derive a minimum brightness temperature of $\sim2.5\,\times 10^6$ K. Optically thick thermal emission of the derived size and temperature would result in a 2--10 keV flux in excess of $10^{-5}$ erg s$^{-1}$ cm$^{-2}$, which is not observed in the data. These values represent very conservative estimates: a smaller region emitting the infrared radiation would result in a higher brightness temperature, which would in turn result in a higher expected X-ray luminosity.
With similar arguments we exclude thermal Bremsstrahlung emission. The existence of an infrared lag is also inconsistent with the  magnetic corona model \cite{merlonietal00}, in which the same population of electrons produces the infrared synchrotron emission and the X-ray Compton emission.

We conclude that the most plausible origin for the observed infrared variability is synchrotron emission from the inner jet. This is confirmed by nearly-simultaneous optical and infrared observations, obtained while the source was in the same low-luminosity state. Those data (Lewis et al., in prep.) show a flat or inverted spectrum (inconsistent with thermal emission from a disc or companion star), and long-timescale ($\sim$minutes) variability stronger in infrared than in optical.

This result is a new, independent strong indication that jet synchrotron emission contributes significantly to the infrared radiation in this source. This is the first time that hard-state, compact jet emission has been securely identified to vary on sub-second timescales in an XB, although
variability on similar dynamical timescales t$_{Dyn}$  (i.e., scaled to mass) had been already observed in Active Galactic Nuclei \cite{schodeletal07}. These data thus represent a further step forward towards a full unification of the accretion/ejection process over a broad range of black-hole
masses.

\subsection{Jet speed and magnetic field}

Our data strongly suggest that the variable infrared emission comes from the jet, although we cannot conclude whether it is optically--thick or --thin synchrotron. The X-ray emission is usually interpreted as Comptonized radiation from energetic plasma in the very inner regions of the accretion flow, although the actual emitting region is still an open issue (either a corona or the base of the jet itself, for a discussion see \cite{markoffetal05,maccarone05}.
Depending on the assumptions we make, we can obtain estimate of different parameters of the jet. 

Namely, if we assume that the infrared emission is thick-synchrotron radiation from the jet, the observed time delay between the infrared and the X-ray variability can give us an upper limit (given the unknown time for the ejection to take place) to the travel time of the variability -- thus presumably the matter -- along the jet. Thus, given a measure of the jet elongation we could estimate the jet speed. Unfortunately such a measure is not available for GX~339-4; however, a jet elongation measurement has been reported from 8.4 GHz observations of another BH XB, Cyg~X-1 \cite{stirlingetal01}. Thus, within the standard model for compact jets \cite{BK79} and assuming that the main physical properties of the jet do not change, we can rescale the jet elongation measured at radio wavelengths in the BH Cyg X-1 down to the infrared wavelengths, obtaining a measurement of the distance of the infrared-emitting region in the jet from the black
hole in GX 339-4. With the caveats of the many key needed assumptions (see \cite{casellaetal10} for a full discussion of the method and its underlying assumptions), we obtain a 3.3-$\sigma$ lower limit on the jet speed of $\Gamma > 2$.

We conclude that, if the infrared synchrotron emission is optically-thick, these data suggest that the jets from accreting stellar-mass BHs are at least mildly relativistic, also in their common low/hard state. If, as is widely suggested, the jet speed corresponds to the escape speed at the launch point, this might imply that the jet is launched from a region very close to the black hole itself.

If on the other hand both the infrared and the X-ray emission are from the very base of the jet, arising from thin-synchrotron radiation, then the above calculation does not hold anymore. Within such an assumption (we refer the reader again to \cite{casellaetal10} for a full discussion of the caveats), the observed time delay between the infrared and the X-ray variability would represent the cooling time of the emitting electron population, from which we obtain an estimate for the magnetic field intensity in the jet of B$\sim 10^4$ Gauss.

\section{The future}

We have detected for the first time fast (sub-second) infrared variability from a jet in an X-ray binary, discovering a clear correlation with the known X-ray variability. We have shown that, within a (large) number of assumptions, this type of data allow us to put quantitative constraints to the jet speed and internal magnetic field. Clearly the obtained estimates have several caveats, or at least large uncertainties. Nevertheless, the potential of the method is revealed, as it offers for the first time the possibility to track the accreting matter from the inflow out in and along the outflow.

Future monitoring observations with this technique will allow to refine these measurements, studying the relative dependency of these quantities with the varying accretion rate or total luminosity. In particular, observations performed simultaneously in (mid-)infrared, X-rays and optical at high time resolution will allow to securely disentangle the different varying components, and possibly separate emission from different regions along the jets. Observations in the hard state at different luminosities will provide information on the evolution of the jet characteristics (many of the uncertainties on the estimates reported above will not affect relative measurements), as for example possible jet acceleration. Similarly, observations in different states, possibly tracking the full spectral evolution during an outburst, will provide information on how and if the jet switches off (or on) during spectral transitions to (or out of) the soft state. Finally, similar observations of neutron-star X-ray binaries will allow to study how the jet properties depend on the type of accreting compact object, possibly unveiling the role of the event horizon and/or ergosphere in the jet-launching mechanism.

The scheduling of such observations is at present severely complicated from the scarcity of permanently mounted fast photometers (needed given the transient nature of the studied objects), and from the lack of {\it multi-wavelength} ones. Nevertheless, some observations have been successfully performed (thanks to the efforts of planners and observers of several facilities, including RXTE, VLT, ULTRACAM and SPITZER), and the analysis is in progress, while others have been approved and will be performed in the following months.

The increased statistics that will be available with the upcoming Extremely Large Telescopes will allow to apply these methods to a large number of sources, otherwise too faint to be observed with the telescopes operating today, eventually offering the possibility to perform population-statistics studies.


\end{document}